\documentclass[10pt]{iopart}

\usepackage{iopams}
\usepackage[square,compress,numbers]{natbib}

\usepackage[utf8]{inputenc} 
\usepackage[T1]{fontenc}    
\usepackage{array}
\usepackage{hyperref}       
\usepackage{url}            
\usepackage{booktabs}       
\usepackage{amsfonts}       
\usepackage{nicefrac}       
\usepackage{microtype}      
\usepackage{lipsum}
\usepackage{multirow}
\usepackage{graphicx}
\usepackage{lineno}

\begin{document}

\title{Lightweight Jet Reconstruction and Identification as an Object Detection Task}
\author{Adrian Alan Pol$^{1,2}$, Thea Aarrestad$^1$, Ekaterina Govorkova$^1$, Roi Halily$^3$, Anat Klempner$^3$, Tal Kopetz$^3$, Vladimir Loncar$^{1,4}$, Jennifer Ngadiuba$^5$, Maurizio Pierini$^1$, Olya Sirkin$^3$, Sioni Summers$^1$}
\address{$^1$European Organization for Nuclear Research (CERN), Geneva, Switzerland \\ $^2$Princeton University, New Jersey, United States \\ $^3$CEVA Inc., Herzliya, Israel \\ $^4$Institute of Physics Belgrade, Belgrade, Serbia \\ $^5$Fermi National Accelerator Laboratory, Illinois, United States }
\ead{adrianalan.pol@cern.ch}
\vspace{10pt}
\begin{indented}
\item[]\today
\end{indented}

\begin{abstract}
We apply object detection techniques based on deep convolutional blocks to end-to-end jet identification and reconstruction tasks encountered at the CERN Large Hadron Collider (LHC). Collision events produced at the LHC and represented as an image composed of calorimeter and tracker cells are given as an input to a Single Shot Detection network. The algorithm, named PFJet-SSD performs simultaneous localization, classification and regression tasks to cluster jets and reconstruct their features. This all-in-one single feed-forward pass gives advantages in terms of execution time and an improved accuracy w.r.t. traditional rule-based methods. A further gain is obtained from network slimming, homogeneous quantization, and optimized runtime for meeting memory and latency constraints of a typical real-time processing environment. We experiment with 8-bit and ternary quantization, benchmarking their accuracy and inference latency against a single-precision floating-point. We show that the ternary network closely matches the performance of its full-precision equivalent and outperforms the state-of-the-art rule-based algorithm. Finally, we report the inference latency on different hardware platforms and discuss future applications.
\end{abstract}

\vspace{2pc}


\maketitle 

\section{Introduction}\label{Section:Introduction}
The world’s largest and most powerful particle accelerator, the CERN Large Hadron Collider (LHC)~\cite{lhc1995large}, operates at a nominal collision rate of $40$~MHz. Due to storage constraints and technological limitations (e.g. fast enough read-out electronics), the volume of recorded data must be significantly reduced by the experiments operating around the accelerator ring. To this purpose, a set of algorithms collectively referred to as the {\em trigger system} are typically used to filter the incoming data stream. Trigger algorithms are designed to reduce the rate of recorded collision {\em events} (e.g., the collection of sensor readouts at each bunch crossing) rate while preserving the physics reach of the experiments. For example, at the Compact Muon Solenoid (CMS) experiment, the trigger system~\cite{khachatryan2017cms, cms2016cms} is structured in two stages using increasingly complex information and more refined algorithms:
\begin{itemize}
    \item The Level 1 (L1) Trigger, implemented on custom-designed electronics; reduces the 40 MHz input to a $100$~kHz rate in $<10~\mu$s.
    \item High Level Trigger (HLT), a collision reconstruction software running on a computer farm; scales the $100$~kHz rate output of L1 Trigger to $1$~kHz in $<150$~ms.
\end{itemize}
With the planned LHC high-luminosity upgrade~\cite{apollinari2017high}, the number of simultaneous collisions per event will surge approximately four-fold. The latency of legacy reconstruction algorithms will increase by more than the factor of three as they may suffer from execution time scaling worse than linearly. Along with the computing infrastructure upgrades, it is worth investigating solutions that could execute many tasks at once, while retaining accuracy and benefiting from the additional speedup offered by parallel computing architectures. Deep neural networks, such as those used for computer vision tasks, are an obvious candidate in this endeavour.

The majority of particles produced in LHC events are unstable and immediately decay to lighter particles. The new particles can decay themselves to others in a so-called decay chain. Such a process terminates when the decay products are stable particles, e.g., charged pions. This collimated shower of particles with adjacent trajectories is called a {\em jet}. Jets are central to many physics studies at the LHC experiments~\cite{butterworth2008jet, skiba2007using, khachatryan2014search, aad2015search}. In particular, a successful physics program requires aggregating particles into jets ({\em jet clustering}), an accurate determination of the jet momentum ({\em momentum measurement}) and the identification of which particle kind started the shower ({\em jet tagging})~\cite{adams2015towards, abdesselam2011boosted, altheimer2012jet, altheimer2014boosted}. 

In this work, we show how jet clustering, momentum measurement, and tagging could all be handled simultaneously on parallel computing architectures. Besides the practical advantages of our approach, one could benefit from multitask learning when accomplishing more tasks at once~\cite{caruana1997multitask}. For instance, a classifier and a regression running at once can learn that calibration constants depend on the nature of the jet, an issue which is now handled with ad-hoc post-processing~\cite{Sirunyan:2019wwa}, i.e. when factorizing the reconstruction problem to energy regression and tagging the overall performance may drop for both. Our main contributions are as follows:
\begin{itemize}
    \item We introduce the {\bf PFJet-SSD} algorithm to perform localization, classification and additional regression tasks on jets in a single feed-forward pass (concurrently, or {\em single-shot}). We combine ideas from different fields of deep learning, i.e. object detection, attention mechanisms, network slimming and quantization.
    \item We report acceleration on different computing architectures.
    \item We generate and publicly share a dataset of simulated LHC collisions, pre-processed to be suited for computer vision applications similar to those discussed in this work, as well as for point-cloud end-to-end reconstruction. The dataset is available on Zenodo~\cite{dataset} and it is accompanied by annotated jet labels, to be used as ground truth during training.
\end{itemize}
The dataset, instructions, and code to fully reproduce our results are available at 
\url{https://github.com/AdrianAlan/PFJet-SSD}.

The remainder of this paper is structured as follows. In Section~\ref{Section:Techniques} we review the key building blocks for this work, i.e. jet images, single-shot detection, attention mechanisms, and efficient model design. In Section~\ref{Section:Methodology} we introduce the PFJet-SSD model and its quantized variants. In Section~\ref{Section:Experiments} we describe the dataset and the training procedure. Finally, in Sections~\ref{Section:Results} and ~\ref{Section:Conclusions} we discuss the results and future directions, respectively.

\section{Techniques}\label{Section:Techniques}
In this section, we review the background for this work, i.e. jet images, single-shot detection, attention mechanism and designing efficient inference networks with pruning and quantization.

\subsection{Jet Images}
Traditional approaches to jet tagging rely on features designed by experts that detect characteristic energy deposit patterns~\cite{plehn2010stop, larkoski2014soft, thaler2011identifying, larkoski2013energy, krohn2010jet, ellis2010recombination, dasgupta2013towards, dasgupta2013jet, dasgupta2015jet}.
In recent years, several studies applied computer vision for event reconstruction at particle colliders, e.g.~\cite{cogan2015jet, almeida2015playing, baldi2016jet, de2016jet, guest2016jet, de2017learning, pearkes2017jet, kasieczka2017deep, komiske2017deep, barnard2017parton, macaluso2018pulling, butter2018deep, lin2018boosting, kasieczka2019machine, bhimji2018deep, nguyen2019topology}. This was obtained by projecting the lower level detector measurements of the emanating particles onto a cylindrical detector and then unwrapping the inner surface of the calorimeter on a rectangle. Such information was further interpreted as an image with calorimeter cells as pixels, i.e. {\em jet images}. This approach was also applied to end-to-end reconstruction, considering not just the individual jet but the whole event~\cite{Andrews:2018nwy,andrews2020end}. Building on these works, we extend the end-to-end reconstruction to include a localization task, merging the jet clustering and classification tasks in a single operation. Centralized computing environments are the only viable options for this: end-to-end approaches require as input a raw data representation, which is not available with reduced analysis data formats. For this reason, we also consider how the model could be compressed to reduce computing footprint, having in mind an approach optimized for a trigger application. 

\subsection{Single Shot Detection}\label{Section:SSD}
Object detection is a fundamental task in computer vision. It is defined as the classification of objects from predefined categories in the image along with their precise spatial locations. The spatial location and extent of an object can be defined coarsely using a bounding box, which is an axis-aligned rectangle tightly bounding the object. Modern object detection focuses on using primarily Convolutional Neural Networks (CNNs) as the building block. Deep learning object detection achieved state-of-the-art results in tasks such as face~\cite{zhang2016joint} or pedestrian detection~\cite{zhang2016faster}. For a general survey on this subject, see~\cite{zou2019object, liu2020deep}.

Deep-learning-based object detection models are typically divided into one-~\cite{redmon2016you, redmon2017yolo9000, lin2017focal, fu2017dssd, zhou2019objects} or two-stage~\cite{girshick2014rich, ren2015faster, girshick2015fast, dai2016r, xu2018deep} detectors. Two-stage detectors generate a sparse set of regions with a high probability of an object being present first (region proposals), followed by a simple classification step. This two-step process is inefficient for real-time applications, due to task serialization. Single-step approaches classify and regress object locations concurrently (in a single feed-forward pass) and as such tend to achieve lower accuracy than two-stage detectors but are simpler and significantly more latency and memory efficient, hence having greater applicability to online problems.

The Single-Shot Multibox Detector (SSD)~\cite{liu2016ssd}, is a simple one-stage, anchor-based detector. First, a set of default regions in an image with a fixed shape and size is predefined to discretize the output space of bounding boxes, called anchors. These anchors have a diverse set of shapes to detect objects with different dimensions, i.e multiple scales and aspect ratios. Based on the ground truth, the object locations are matched with the most appropriate anchors to obtain the supervision signal for the anchor estimation. At inference, each anchor is refined by four box coordinates (width, height, x and y offsets) and predicts the categorical probabilities. To avoid a huge number of negative proposals dominating training gradients, hard negative mining is used to train the network, which fixes the foreground and background ratio~\cite{felzenszwalb2009object}\footnote{By background we refer to the areas without objects.}. Alternatively, a focal loss~\cite{lin2017focal} could be used. In this case, the price to pay would be more hyperparameters to tune. The SSD architecture is fully convolutional, with initial layers based on a pre-trained backbone architecture, such as VGG-16~\cite{simonyan2014very}, followed by extra convolutional and pooling layers which progressively decrease image size and thus increase the receptive field. The information in the last layer may be too coarse spatially to allow precise localization and at the same time, detecting large objects in shallow layers is non-optimal without large enough receptive fields. As a countermeasure for this issue, the SSD performs detection over multiple scales by operating on multiple feature maps, i.e. at different depths of the network. Each of these feature maps is responsible for detecting objects according to their receptive field. To detect large objects and increase receptive fields extra convolutional feature maps were added to the backbone architecture. The final prediction is made by merging all detection results from different feature maps followed by a Non-Maximum Suppression (NMS)~\cite{liu2016ssd} step and producing the final detection information. NMS removes duplicate predictions originating from multiple anchors.

\subsection{Attention Mechanisms}
Visual Attention Gates (AGs), e.g.~\cite{jetley2018learn, anderson2018bottom, oktay2018attention}, learn to suppress feature activations in irrelevant regions in an input image without additional supervision. At inference, the gates generate soft region proposals to highlight salient features useful for a specific task. Recently, the performance of deep CNNs on visual tasks was improved with scale-aware~\cite{li2019scale, yi2019assd}, spatial-aware~\cite{woo2018cbam, fu2019dual} and channel-wise~\cite{hu2018squeeze, chen20182} attention. On the contrary, most of the attention modules inevitably increase model complexity. Efficient Channel Attention (ECA) gate~\cite{wang2020ecanet} is a soft attention mechanism that addresses this issue. It avoids dimensionality reduction and captures cross-channel interaction efficiently. ECA gate $\omega$ is given by $\omega=\sigma({\textbf W} \odot g(y)),$ where $y \in \mathbb{R}^C$ is the feature map activation with channels $C$, $g$ is channel-wise global average pooling, $\sigma$ is the Sigmoid function and $\textbf W$ is a weight tensor of a $1D$ convolution of filter size $k$.
 
\subsection{Quantization}
Optimizing deep neural networks for efficient inference is an essential task in modern machine learning pipelines due to limitations presented by edge devices. Models should provide high accuracy with a minimum of computing time and resources. Apart from accelerating inference online, e.g. through parallelization or hardware optimizations, models can be optimized offline, through compression~\cite{han2015deep}.

Network compression~\cite{cheng2017survey} is a common technique to reduce the number of operations and model size, energy consumption, and over-training of deep neural networks. As neural network synapses and neurons can be redundant, compression techniques attempt to reduce the total number of them, effectively reducing multipliers. Several approaches have been successfully deployed without much loss in accuracy, including selective removal of parameters based on a particular ranking and regularization, i.e. parameter pruning~\cite{lecun1989optimal, louizos2017learning,gordon2018morphnet}, compact network architectures~\cite{howard2017mobilenets, iandola2016squeezenet, cohen2016group}, and reducing the precision of operations and operands, i.e. quantization~\cite{courbariaux2015binaryconnect, courbariaux2016binarized, zhou2016dorefa, rastegari2016xnor, hubara2017quantized, zhu2016trained, lee2017lognet, cai2017deep}.

It has been observed that reducing the precision of the calculations, i.e. weights and biases, has little impact on performance compared to speedup and resource usage gains. This includes moving away from $32$-bit floating-point calculations (or {\em full-precision}, FP) to fixed points, reducing bit-width and weight sharing. An example of a very aggressive strategy is reducing weight precision to ternary values restricted to $\{-1,0,1\}$ only, called Ternary Weight Network (TWN)~\cite{li2016ternary}. The quantization is performed during training, using a straight-through estimator~\cite{courbariaux2015binaryconnect}, where ternary weights are used during the forward and backward propagation but not during the parameter update. 
To quantize the full precision weights \textbf{W} to ternary ones \textbf{W}$^*$ , TWN uses a threshold value $\Delta$:
\begin{equation*}
\textbf{W$^{*}$} = \left\{  \begin{array}{r@{\quad}cl} 
  +1 & \mathrm{ if } & \textbf{W} > \Delta \\
  0 & \mathrm{ if } & |\textbf{W}| \leq \Delta \\
  -1 & \mathrm{ if } & \textbf{W} < -\Delta
\end{array}\right.
\end{equation*}
with approximated solution $\Delta^*\approx0.7\cdot\mathrm{E}(|\textbf{W}|)$. To make the network perform well, TWN minimizes the Euclidian distance between \textbf{W} and \textbf{W$^{*}$} along a non-negative scaling factor $\alpha$ that can be implemented with per-network, per-layer or per-channel granularity, transforming the weights to $\alpha\textbf{W}^{*}$. For any $\Delta$ the optimal $\alpha$ is computed as: $\alpha_\Delta^*=\frac{1}{\textbf {I}_\Delta}\sum_{i \in \textbf{I}_\Delta}|\textbf{W}_i|,$ where $\textbf{I}_{\Delta}=\{i \big| |\textbf{W}_i|>\Delta\}$ and $|\textbf{I}_\Delta|$ denotes number of elements in $\textbf{I}_\Delta$.

\begin{figure}[ht]
  \centering
  \includegraphics[width=\linewidth]{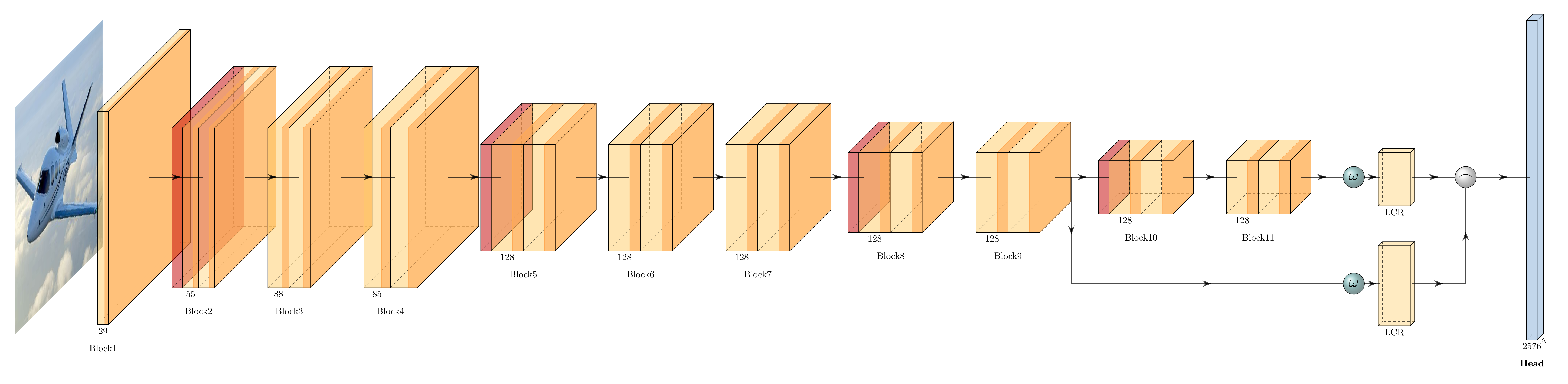}
  \caption{PFJet-SSD architecture. The convolution block ($3 \times 3$ convolution followed by batch normalization and PReLU activation) is in yellow, the average pooling ($2 \times 2$ kernel) is in red, the detection head which is the output layer is in blue. $\omega$ is the attention module and $\frown$ is the concatenation. The numbers indicate the number of output channels in each block. \texttt{Block10} and \texttt{Block11} are removed at inference.}
  \label{Figure:PFJet-SSD-Architecture}
\end{figure}

\section{Methodology}\label{Section:Methodology}
The PFJet-SSD architecture is shown in Figure~\ref{Figure:PFJet-SSD-Architecture}. We modify the original SSD architecture~\cite{liu2016ssd} and Jet-SSD architecture proposed in~\cite{pol2021jet}. Having in mind an HLT application with a typical latency of $\approx 150$~ms, we extend the event image representation to include the information from the charged-particle reconstruction. We do so by adding a {\em tracker} channel to the image, in front of the calorimeter channels already introduced in~\cite{pol2021jet}. We use a lightweight MobileNet architecture~\cite{howard2017mobilenets} as a backbone for our detector which replaces the convolution operation with a combination of depthwise and pointwise versions. Each convolution is followed by a batch normalization~\cite{ioffe2015batch, sari2020does} and parametric rectified linear unit (PReLU)~\cite{he2015delving} activation layers. We use the \texttt{AveragePool} layer to decrease the size of the feature map. The extra convolutional layers proposed by the original SSD do not contribute to accurate detection (recall the remark about the increasing receptive field from Section~\ref{Section:SSD}). This is due to the size of the jets. As done in~\cite{pol2021jet} we remove these layers already at the training time. Retaining the deeper layers of the backbone, i.e. \texttt{Block10} and \texttt{Block11}, does not show improvements at inference but is necessary during training due to additional signals during back-propagation. Hence, these deeper layers are only purged after training, i.e. the concatenation layer ignores them only at inference.

The detection head output corresponds to jet class, localization ($\eta$ and $\phi$ offsets) and $p_T$ value. One might easily extend this output to include jet mass regression as well (we left this out for simplicity). For localization, we regress only the centre of the jet, as we can determine its size from its class. This allows us to set only one scale and one aspect ratio for anchors in each feature map which reduces the complexity of the network.

Finally, we add two new modules to the network. First, the initial convolutional layer is now followed by spatial dropout~\cite{ghiasi2018dropblock} (with $p=0.1$). Second, we attach the ECA gate~\cite{wang2020ecanet} (with $k=3$) before the Localization Classification Regression (LCR) layer.

We use magnitude pruning~\cite{han2015learning} during training to find the optimal allocation of resources between layers. Unstructured pruning generality leads to a higher compression rate and/or higher accuracy when compared to the structured version, but it requires special software or hardware accelerators to fully benefit from it. Since the outcome of an unstructured pruning is a sparse tensor, one needs a dedicated way to handle sparse memory access on hardware to turn pruning compression into a computational advantage at inference time. We use an alternative, a version of structured pruning that removes whole channels in a convolutional block slimming the network without increasing sparsity. We target the hardware implementation that benefits from fusing batch normalization and convolution parameters at runtime. Doing so, the target filter weights $\textbf{W}$ of block $l$ are $\textbf{W}_l=\gamma_l\textbf{W}_l^{\texttt{conv}}$, where the $\textbf{W}^{\texttt{conv}}$ are the weights of the convolution and $\gamma$ is the scale parameter of the affine transformation of the subsequent batch normalization layer. We thus add a regularizer that pushes the influence of filters down through batch normalization $\gamma$ L1 penalty, similarly to~\cite{gordon2018morphnet,liu2017learning}. We scale this penalty based on the number of operations $\mathcal{O}$ in each layer. The sparsifying regularizer $\texttt{G}(\theta)$ is calculated as $\texttt{G}(\theta)=\sum_l|\gamma_l|\mathcal{O}_l$. We mark channels to prune based on the $\gamma$ distribution in each layer, using the rule: $|\gamma_l|<\mu(|\gamma_l|)-\sigma(|\gamma_l|)$. When this rule is not sufficient to remove the specified number of channels we simply select the remaining ones based on ascending magnitudes of $\gamma$.

Also during training, we quantize the network to homogeneous $8$-bit fixed point precision for both weights and activations and $2$-bit TWN with layer- and channel-dependent scaling factors. For the latter, we experimented with a grace period of frozen quantization for which the $\Delta$ and $\alpha$ parameters remain unchanged. Training TWN in this manner may offer greater stability, i.e. weights have time to adjust to new parameters, but in our case, the final results did not improve.

\section{Experiments}\label{Section:Experiments}
In this section, we review the experimental dataset and training procedure used for the experiments.

\begin{figure}[ht]
  \centering
  \includegraphics[width=\linewidth]{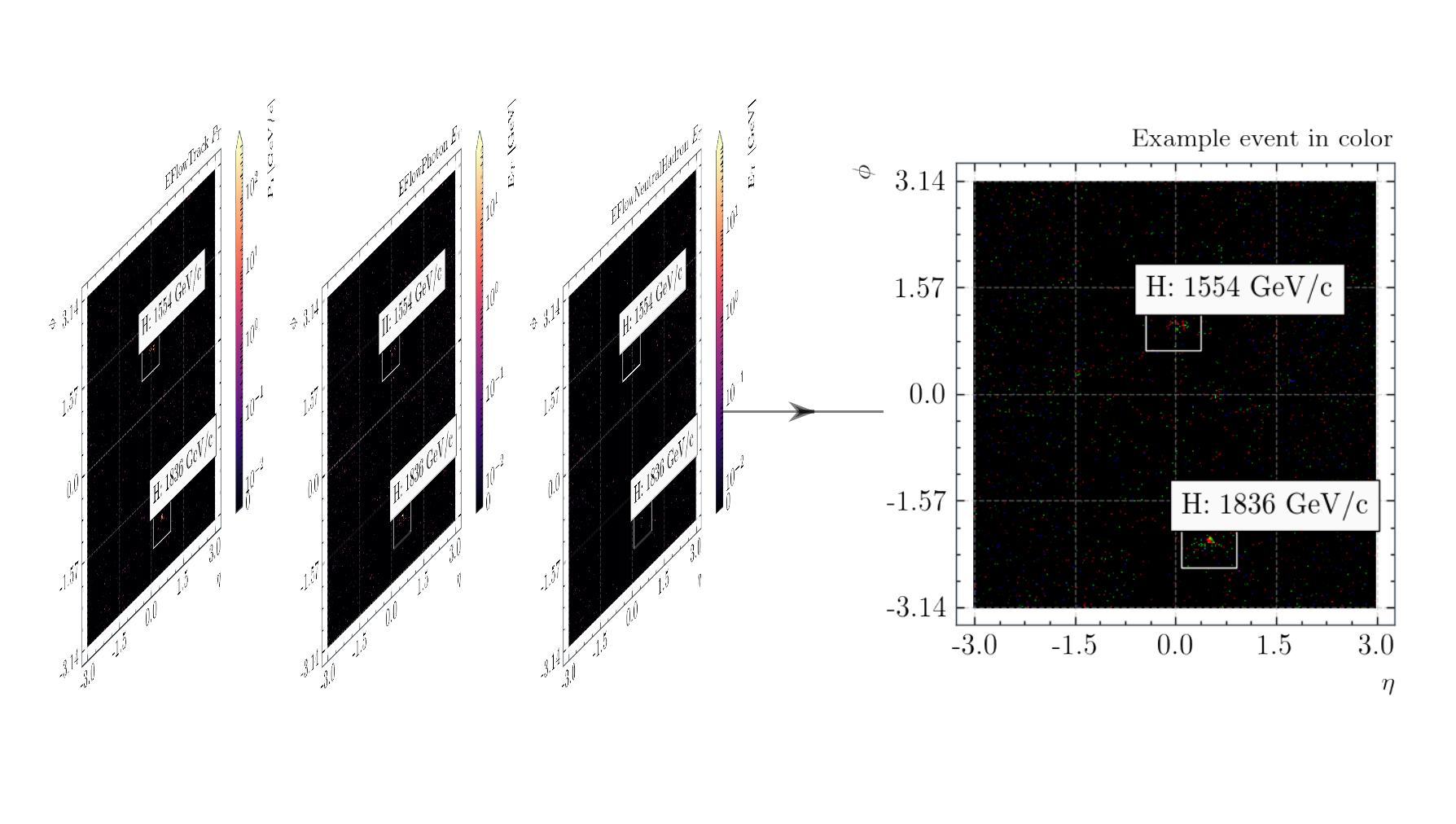}
  \caption{An example input to the PFJet-SSD network: tracker information and energy deposits in CMS ECAL and HCAL translated to a two-dimensional image. The white bounding boxes correspond to ground truth with target label and momentum.}
  \label{Figure:Example-Event}
\end{figure}

\paragraph{Dataset}
The input dataset  consists of $13$~TeV proton-proton collision events, in which Randall-Sundrum gravitons decay to $b \bar b$, gg, qq, HH, WW, ZZ, or $t \bar t$ final states. The choice of this particular process is motivated by the possibility of creating well-defined jet pairs belonging to specific jet classes and with the same kinematic properties across classes. In addition to the hard collision, parasitic {\it pileup} collisions are also simulated, overlapping minimum bias events. The number of pileup collisions is sampled from a Poisson distribution.

The detector effects and hadronization have an important effect on a jet substructure. Events are generated with Pythia~\cite{sjostrand2008brief}. We use the CMS Delphes~\cite{de2014delphes} description to mimic the effect of detector reconstruction. The core of the CMS detector is a multi-layer silicon tracking device, operating in a $4$T magnetic field. Two calorimeter layers surround the tracker: the lead tungstate crystal Electromagnetic Calorimeter (ECAL) is designed to stop particles whose main interaction is electromagnetic (photons and electrons); the brass and scintillator Hadronic Calorimeter (HCAL) is designed to stop hadrons. They give a measurement of the energy of particles (charged and neutrals). Each of them is composed of a barrel and two endcap sections. Forward calorimeters extend the pseudorapidity ($\eta$) coverage provided by the barrel and endcap detectors. A more detailed description of the CMS detector, together with a definition of the coordinate system used and the relevant kinematic variables, can be found in~\cite{collaboration2008cms}.

The calorimeter cells (towers) in the barrel region together with tracker cells are arranged in a fixed discrete space with fine segmentation in $\eta$ and $\phi$, where $\phi$ is the translated azimuthal angle. Before the LHC, jets were usually reconstructed from their calorimeter deposits (known as CaloJet). With the start of the LHC, the CMS particle flow (PF) algorithm~\cite{sirunyan2017particle} demonstrated that the additional information from track reconstruction could increase the accuracy of jet reconstruction. In CMS, this was crucial to compensate for the poor energy resolution of the HCAL. In the long term, this strategy was found to be effective beyond jet momentum measurement, since the angular resolution of the tracking algorithm provided valuable information for jet tagging and substructure algorithms. The PF algorithm for jet reconstruction was eventually adopted also by the ATLAS experiment~\cite{ATLAS:2017ghe}. Taking this as our starting point, we build our event image starting from PFcandidates (as returned by the Delphes PF algorithm), arranging the particles in three groups: charged particles, used to create the tracker channel; photons and electrons, used for the ECAL channel; neutral hadrons, used for the HCAL channel. In a real-life application, one could use the same approach or build the channels from the raw detector hits in the tracker, ECAL, and HCAL. The best approach to follow depends on the context of the application (e.g., online vs offline). We unwrap the cylindrical detector to compose the final image which is formed by translating the calorimeter energy deposits and tracker momentum into pixels using ECAL granularity, which results in $340\times360\times3$ pixel samples. An example is shown in Figure~\ref{Figure:Example-Event}. Some previous studies on jet images implemented data pre-processing steps such as translation, rotation, re-pixelation, or inversion. However, in our study, we only limit the input to $\eta \in (-3, 3)$ and standardize pixel intensities.

\begin{figure}[ht!]
  \centering
  \includegraphics[height=8cm,keepaspectratio]{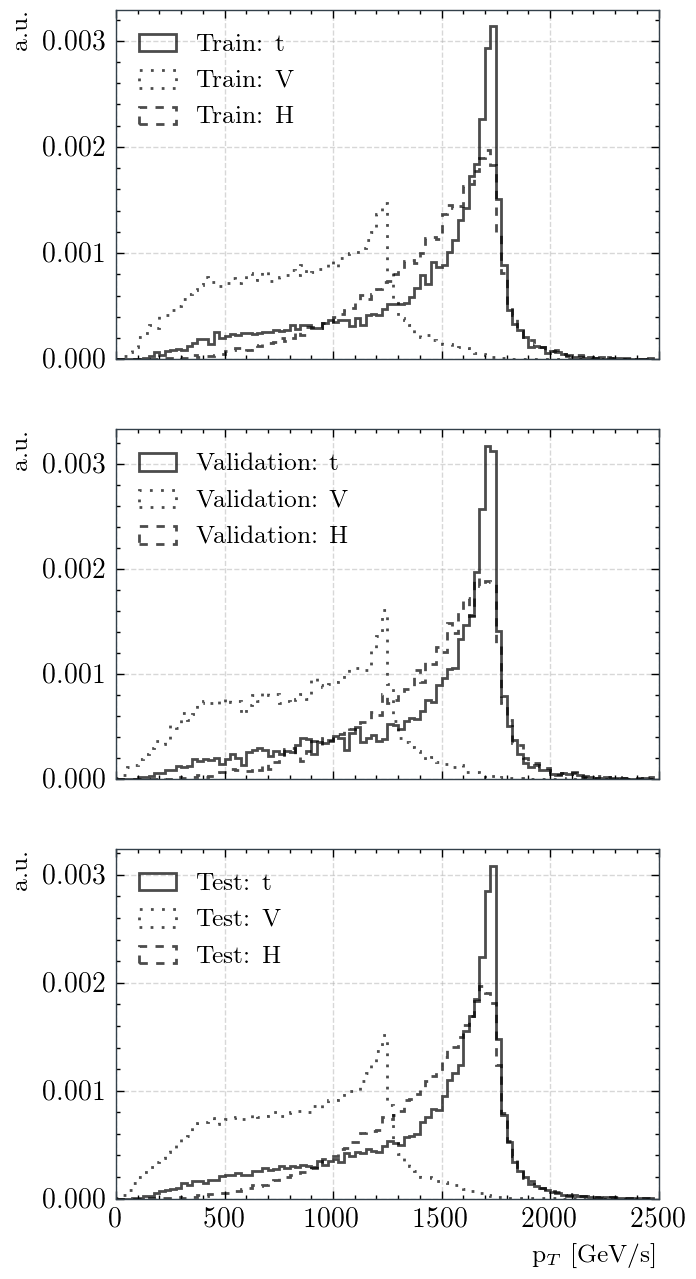}
  \includegraphics[height=8cm,keepaspectratio]{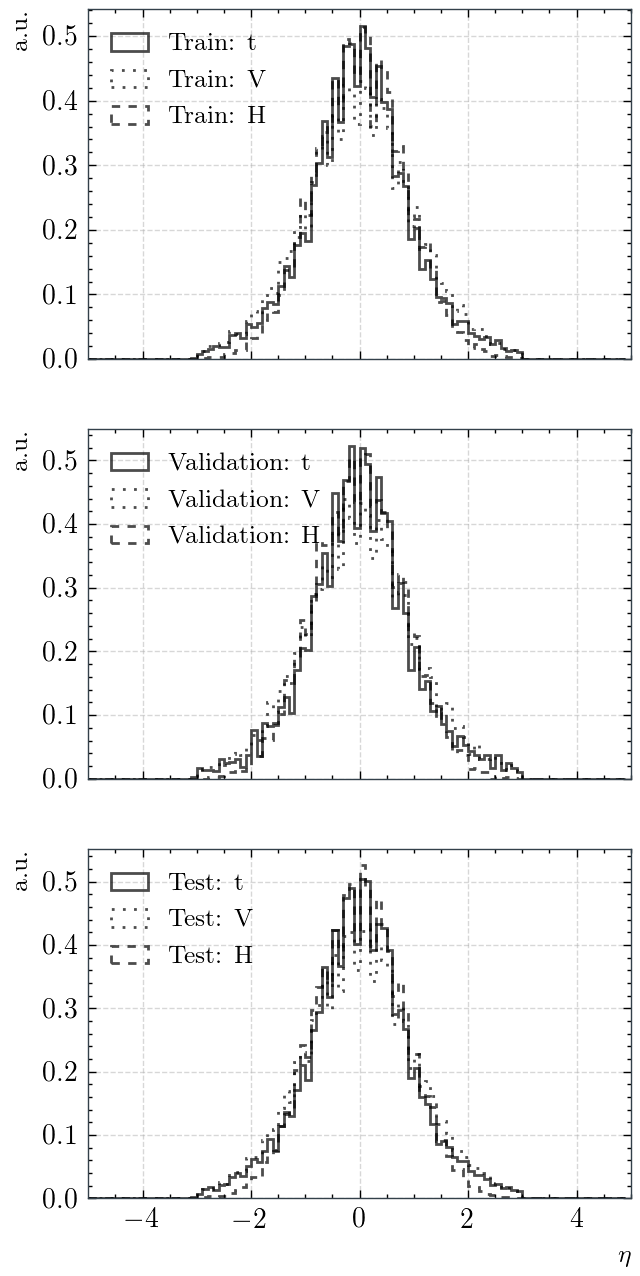}
  \includegraphics[height=8cm,keepaspectratio]{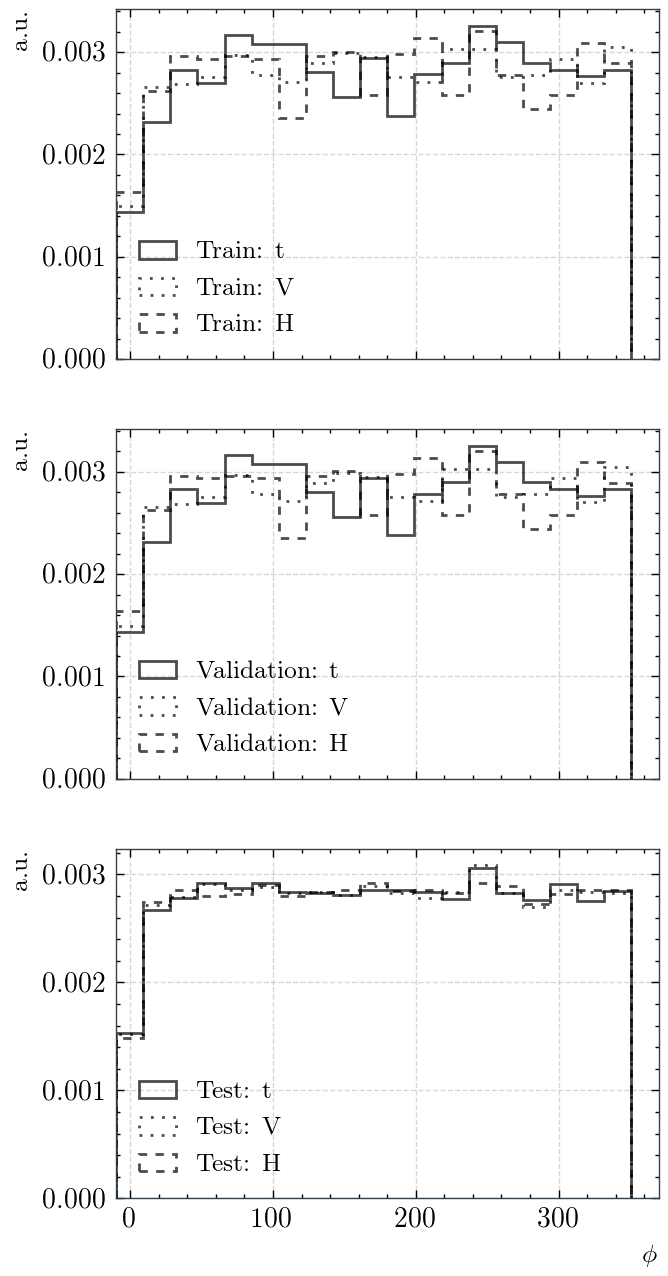}
  \caption{Dataset profile as a function of $p_T$ (left), $\eta$ (middle) and $\phi$ (right).}
  \label{Figure:Dataset-Profile}
\end{figure}

Jet labels are obtained using generator-level information. We assign the jet $\eta$, $\phi$ and $p_T$ measurements to the properties of the same particle. The minimum jet $p_T$ in the dataset is $7$~GeV. Details on the dataset profile are given in Table~\ref{Table:Jet-Statistics} which describes the jet statistics across datasets. Figure~\ref{Figure:Dataset-Profile} shows the $p_T$, $\eta$ and $\phi$ distributions.

\begin{table}[ht!]
    \begin{center}
        \begin{tabular}{| l || c | c | c |}
            \hline & Train & Validation & Test \\ \hline
            t & 59388 & 23802 & 59392 \\ \hline
            V & 118701 & 47493 & 118832 \\ \hline
            H & 59967 & 23997 & 59978 \\ \hline
        \end{tabular}
    \end{center}
    \caption{Number of samples in the datasets.}
    \label{Table:Jet-Statistics}
\end{table}

\paragraph{Training Procedure}
The PFJet-SSD network is implemented on NVidia Tesla GPUs using PyTorch~\cite{NEURIPS2019_9015}. For training, we use stochastic gradient descent with an initial learning rate of $10^{-3}$ with momentum set to $0.9$ and weight decay to $0.0005$. We train the network for $100$ epochs with a batch size of $25$, decreasing the learning rate by a factor of $2$ after every $10$ epochs after the $20^{th}$ epoch. We use $90$k and $36$k samples for training and validation, respectively. The training is performed in mixed-precision to speed up computation and distributed across $3$ GPUs. Thus, we replace the standard batch normalization layer with the \texttt{SyncBatchNorm} layer provided by PyTorch to synchronize statistics across the machines while training.

We minimize the following cost function: $$\mathcal{L_\textbf{SSD}} = \mathcal{L}_c + \mathcal{L}_l + \mathcal{L}_r,$$ where the $\mathcal{L}_c$ is the classification loss, the $\mathcal{L}_l$ is the localization loss, the $\mathcal{L}_r$ is the regression loss. We use cross-entropy with smooth labels ($\alpha=0.1$) for classification, and Huber loss~\cite{girshick2015fast} ($\delta=1$) for localization and regression.

A common challenge when training object detection models from scratch is the insufficient amount of training data which may lead to overfitting \footnote{That is not a problem in our case as we can generate more events with low cost}. Thus it is common to see practitioners pre-loading weights from pre-trained classification models on the real-world ImageNet~\cite{5206848} dataset. We found that such a procedure slows down our learning as the real-world images have little relation to our calorimeter images. The full precision network (FPN) can learn faster by using Xavier uniform initialization~\cite{pmlr-v9-glorot10a} (which helps with the sparsity of the input). We also augment the training dataset by random flips along $\eta$ and $\phi$ dimensions, which we find to greatly stabilize the training. We did not experiment with other augmentation techniques such as changing brightness, contrast, saturation and hue as jets are not invariant to such transitions. The experiments on other commonly used techniques such as Mix-Up~\cite{zhang2017mixup} or Mosaic~\cite{bochkovskiy2020yolov4} yield subpar results, again. This is likely because of the different nature of our input.

We perform five steps of iterative pruning, each with $20$ epochs of retraining a gradually decreasing number of channels in each block. We then retrain the network for the last time for $100$ epochs. We found out that pre-loading FPN weights when training the quantized versions, i.e. TWN and $8$-bit fixed-precision (INT8) network, greatly speeds up convergence.

\section{Results}\label{Section:Results}
In this section, we present the detection and latency performance of PFJet-SSD.

\begin{figure}[ht]
  \centering
  \includegraphics[width=\linewidth]{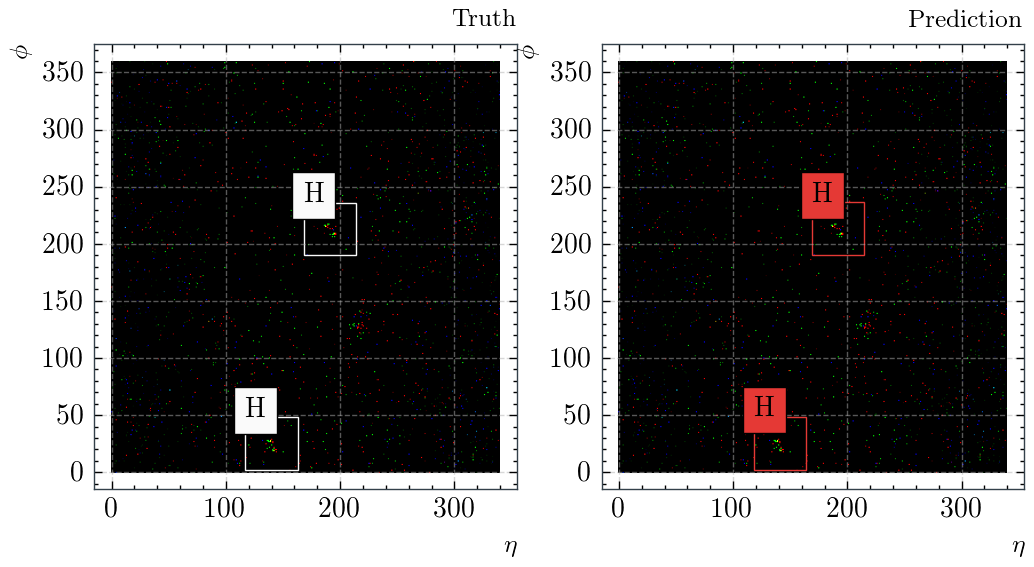}
  \includegraphics[width=\linewidth]{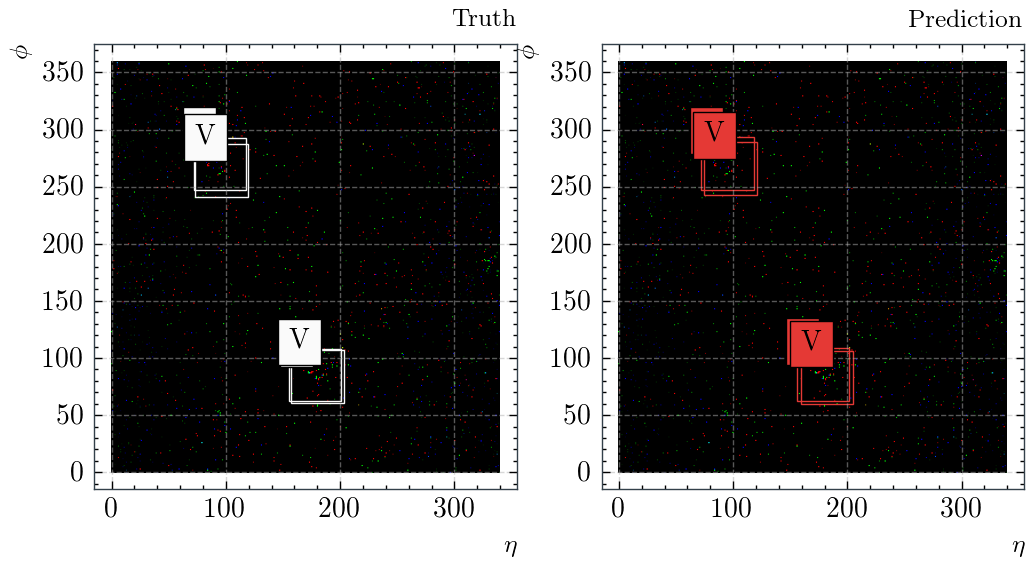}
  \caption{Two examples of the PFJet-SSD at inference for two events with the input image and highlighted true labels (left) and predicted bounding boxes (right). The overlapping boxes in the second event correspond to the $t \rightarrow bW$ decay where two jets, t and W, are very close.}
  \label{Figure:Example-Inference}
\end{figure}

\begin{figure}[ht]
  \centering
  \includegraphics[width=.5\linewidth]{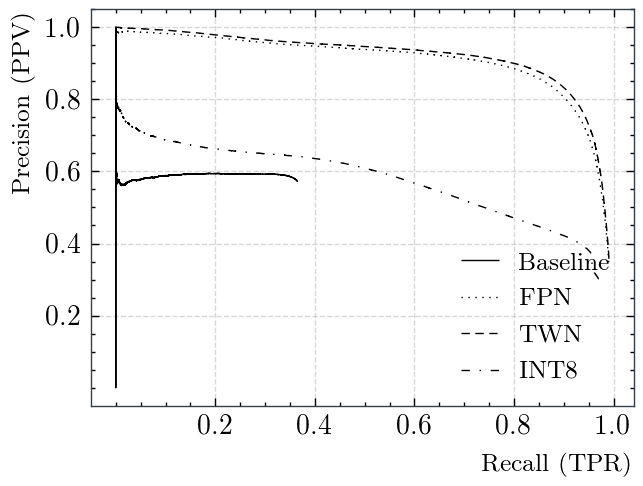}
  \includegraphics[width=.5\linewidth]{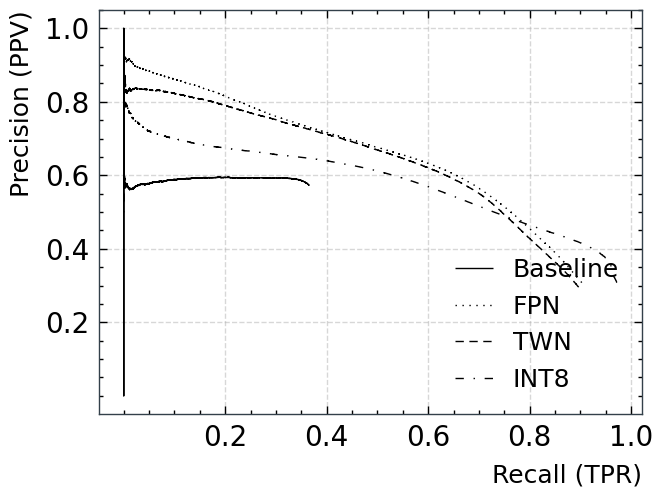}
  \includegraphics[width=.5\linewidth]{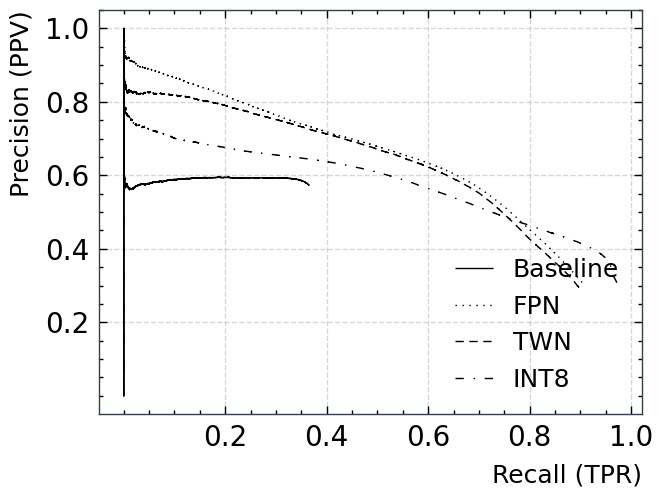}
  \caption{Precision-Recall curves for the baseline algorithm and three flavours of the PFJet-SSD model. The inference is performed on original, non-mixed samples (left), two mixed events in superposition (centre), three  mixed events in superposition (right).}
  \label{Figure:Precision-Recall-Curve}
\end{figure}

\begin{table}[ht]
    \begin{center}
        \begin{footnotesize}
            \begin{tabular}{| c | c || c | c | c | c |}
                \hline
                \multicolumn{2}{|c||}{} & \multirow{2}*{Physics Baseline} & \multicolumn{3}{c|}{PFJet-SSD} \\
                \cline{4-6}
                \multicolumn{2}{|c||}{} & & FPN & TWN & INT8 \\
                \hline
                \multicolumn{2}{|c||}{NoP} & N/A & \multicolumn{3}{c|}{111 228} \\
                \hline
                \multicolumn{2}{|c||}{NoOps} & N/A & \multicolumn{3}{c|}{1.095G} \\
                \hline
                \multicolumn{2}{|c||}{W/A} & N/A & 32/32 & 2/32 & 8/8 \\
                \hline
                \multicolumn{2}{|c||}{AP} & .161 & .848 & .857 & .566 \\
                \hline
                \multirow{3}*{top jet} & AP & .420 & .865 & .872 & .473 \\
                \cline{2-6}
                & P@R=.3 & .736 & .985 & .988 & .531 \\
                \cline{2-6}
                & P@R=.5 & .627 & .975 & .980 & .453 \\
                \hline
                \multirow{3}*{W/Z jet} & AP & .245 & .847 & .859 & .629 \\
                \cline{2-6}
                & P@R=.3 & .584 & .944 & .955 & .673 \\
                \cline{2-6}
                & P@R=.5 & N/S & .929 & .943 & .653 \\
                \hline
                \multirow{3}*{H jet} & AP & .107 & .860 & .872 & .335 \\
                \cline{2-6}
                & P@R=.3 & N/S & .992 & .996 & .453 \\
                \cline{2-6}
                & P@R=.5 & N/S & .978 & .986 & .400 \\
                \hline
            \end{tabular}
            \caption{Detection details for the baseline and PFJet-SSD algorithms, reporting the number of parameters (NoP), the number of operations (NoOps), the precision of weights/activations (W/A), average precision (AP) and precision at $0.3$ (P@R=.3) and $0.5$ (P@R=.5) recall. The table does not report parameters and bit precision for the baseline as it is a non-parametric method: not applicable (N/A). The baseline is also unable to reach 0.3 and 0.5 in several cases: no statistics (N/S).}
            \label{Table:Precision-Recall-Details}
        \end{footnotesize}
    \end{center}
\end{table}

\begin{figure}[ht]
  \centering
  \small{t}\\
  \begin{tabular}{cc}
      & \multirow{3}*{\includegraphics[width=0.85\linewidth]{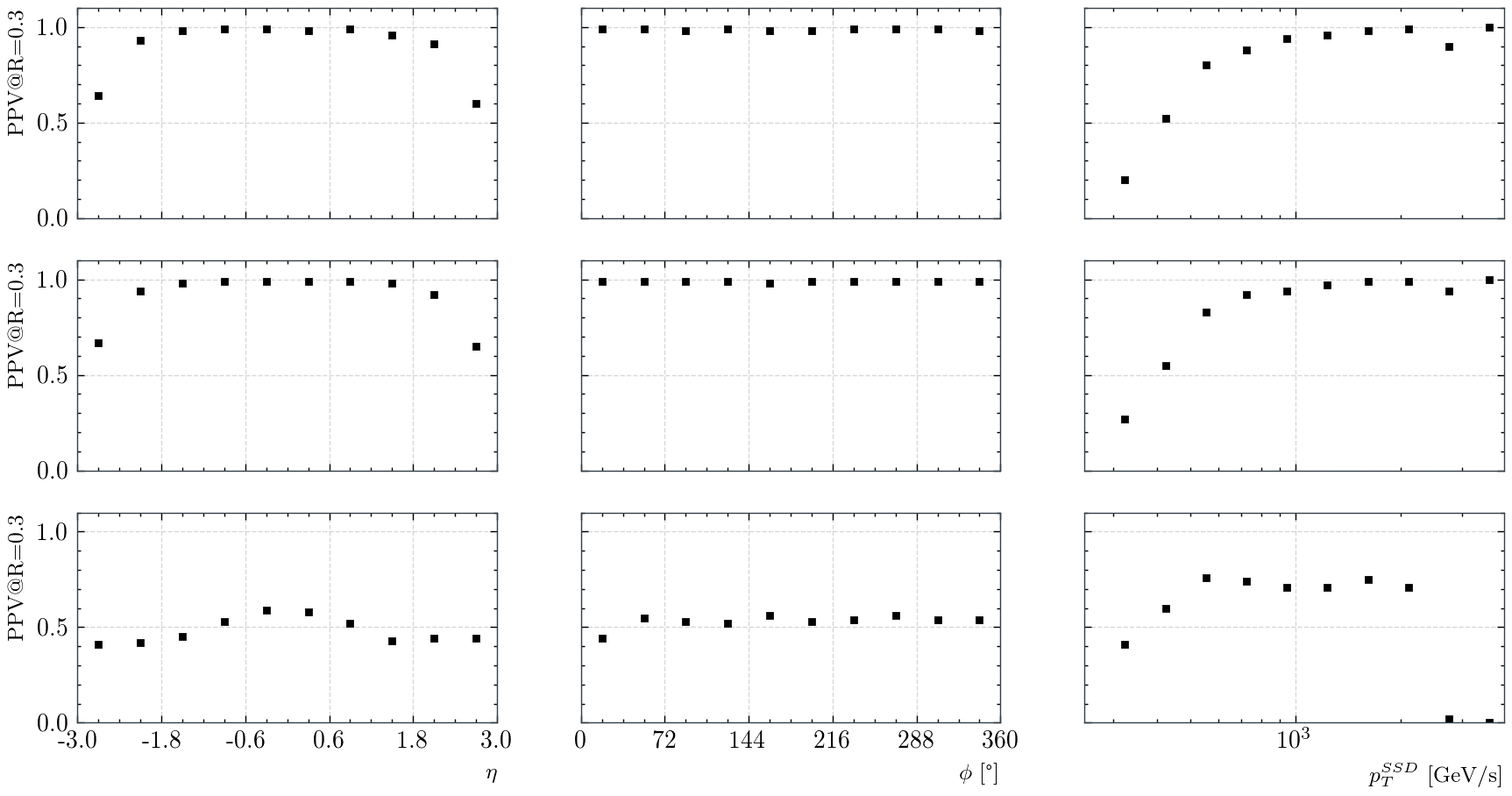}} \\
      \\ FPN \\ \\[3.3em] TWN & \\ \\[3.3em] INT8 & \\[3.3em]
  \end{tabular}

  \small{V}\\
  \begin{tabular}{cc}
      & \multirow{3}*{\includegraphics[width=0.85\linewidth]{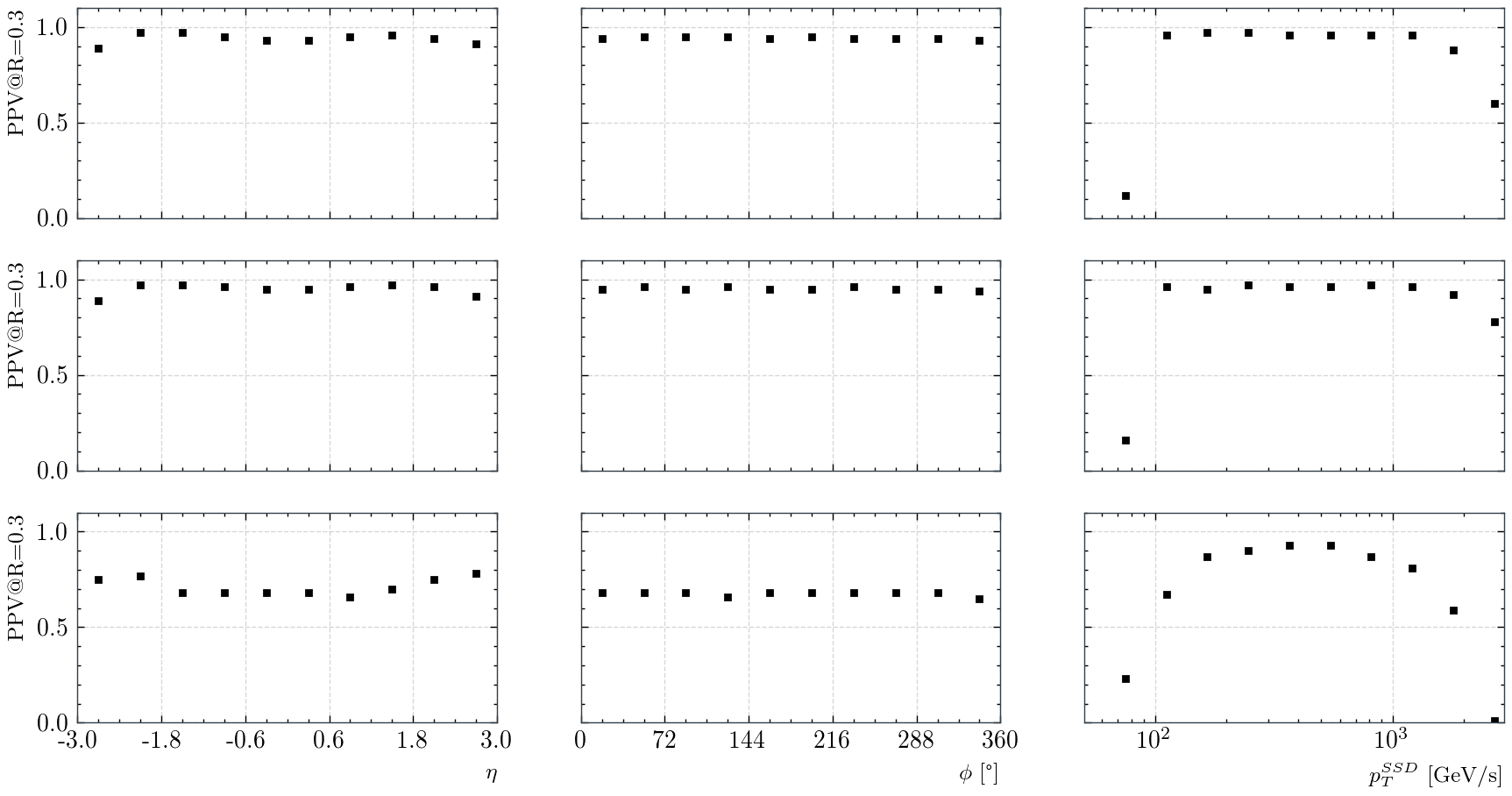}} \\
      \\ FPN \\ \\[3.3em] TWN & \\ \\[3.3em] INT8 & \\[3.3em]
  \end{tabular}

  \small{H}\\
  \begin{tabular}{cc}
      & \multirow{3}*{\includegraphics[width=0.85\linewidth]{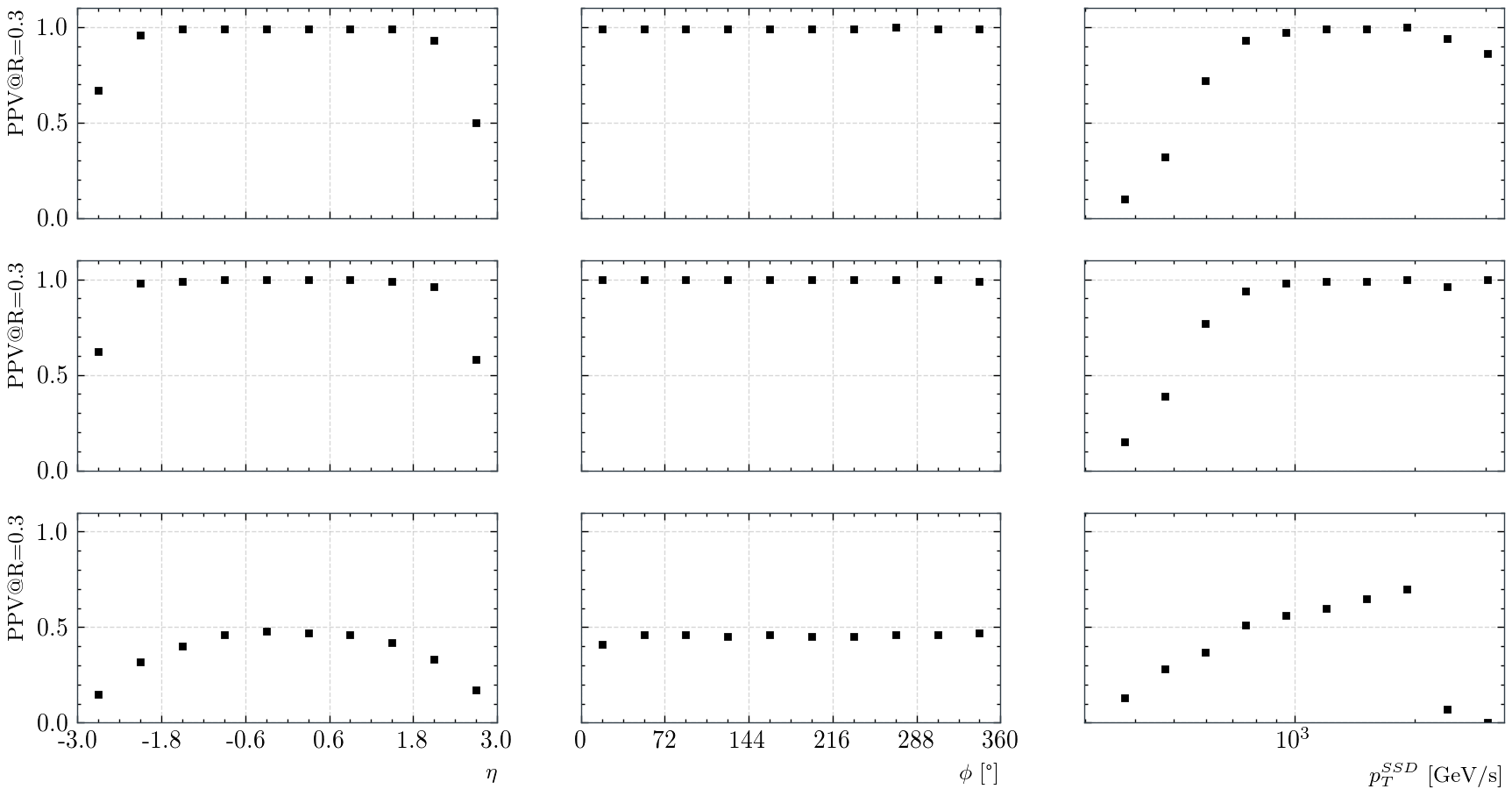}} \\
      \\ FPN \\ \\[3.3em] TWN & \\ \\[3.3em] INT8 & \\[3.3em]
  \end{tabular}

  \caption{Precision at $30\%$ Recall (PPV@R=03(PPV) Recall for top (top), V (centre) and H (bottom) jets as a function of $\eta$ (left), $\phi$ (middle) and $p_T$ (right). For each block of figures (top, V, and Higgs), we show results for the FPN (top), TWN, and INT8 models.\label{Figure:Precision-Details}}
\end{figure}

\begin{figure}[ht]
  \centering
  t \\ \includegraphics[width=\linewidth]{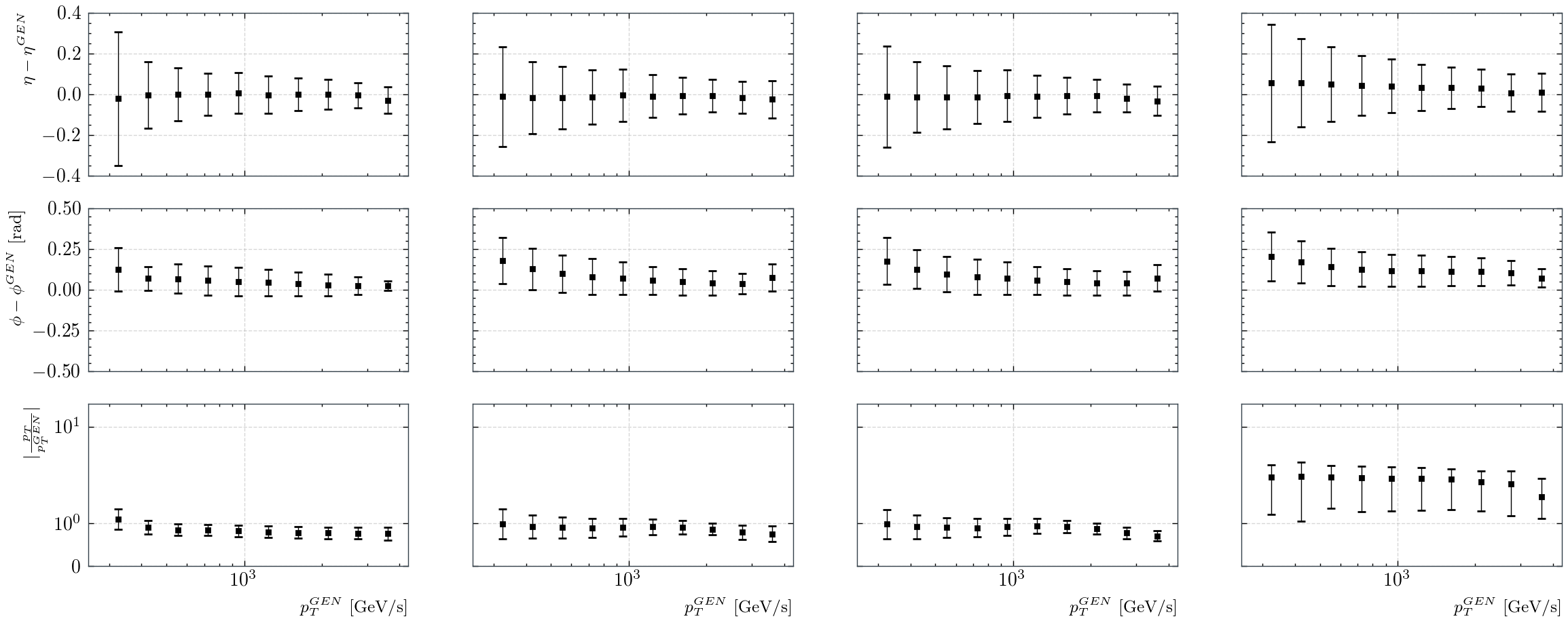}
  
  V \\ \includegraphics[width=\linewidth]{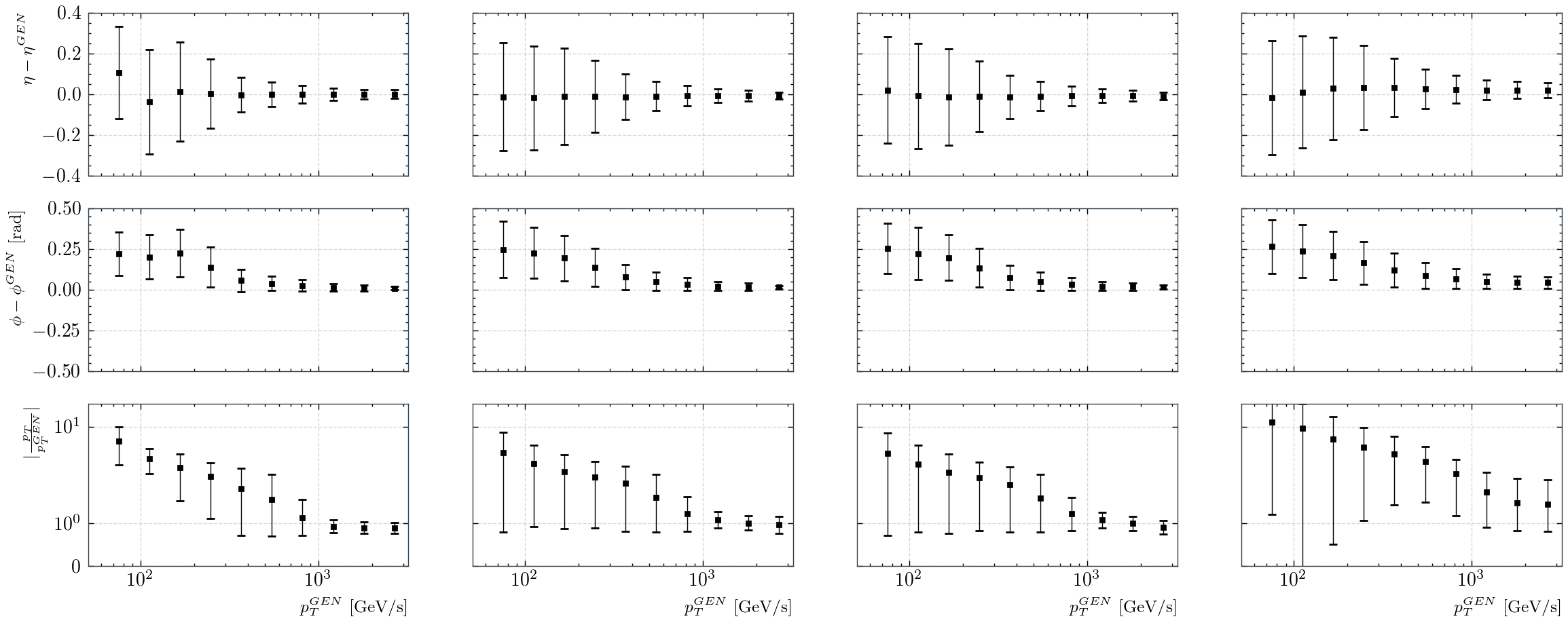}
  
  H \\ \includegraphics[width=\linewidth]{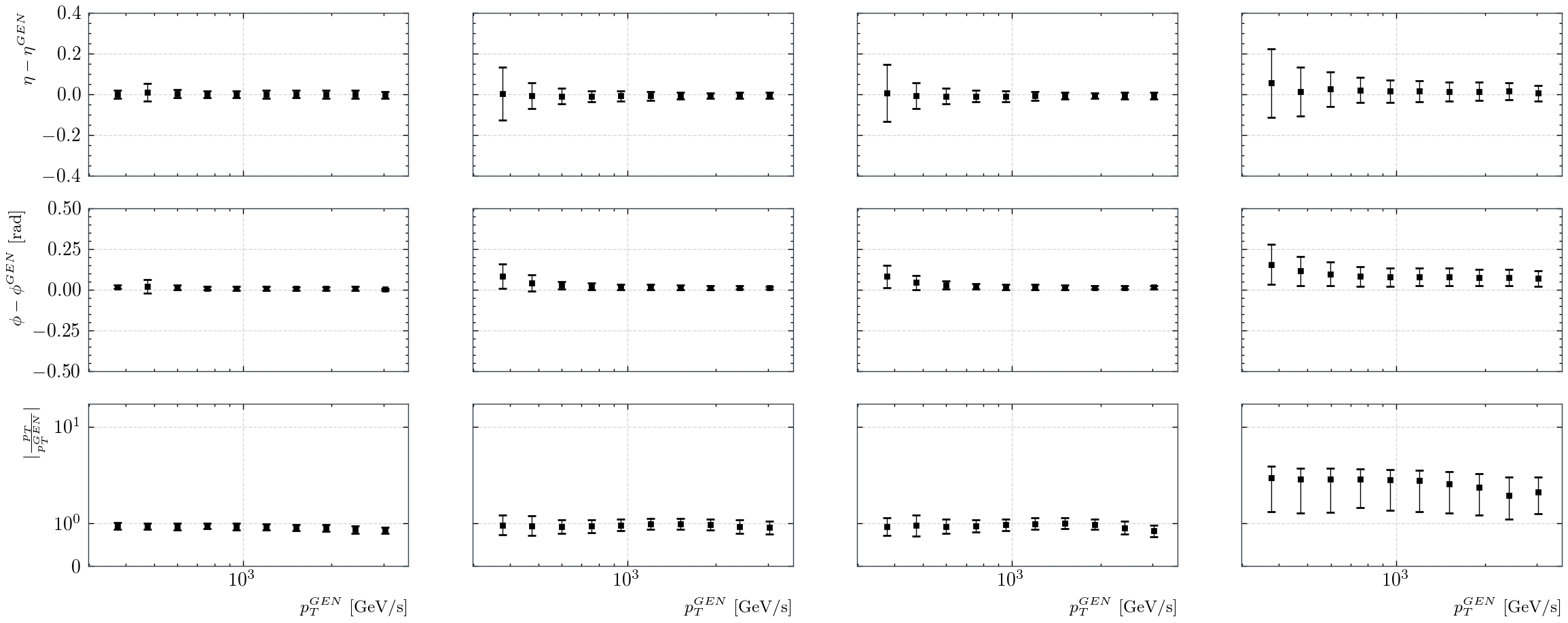}
  \caption{Displacement in $\eta$, $\phi$, and relative $p_T$ regression error (bottom row) for top (top), V (centre) and H (bottom) jets, as a function of generator-level $\eta$ (left), $\phi$ (middle) and $p_T$ (right). For each block of figures (top, V, and Higgs), we show results for the FPN (top), TWN, and INT8 models.\label{Figure:Displacement-Regression-Error}}
\end{figure}

\paragraph{Detection Performance}
As a proof of concept, we investigate the tagging of the top (t), W and Z boson (V) and H boson (H) jet. An example of the PFJet-SSD output is shown in Figure~\ref{Figure:Example-Inference}. PFJet-SSD outputs predicted categorical label, prediction confidence and the centre coordinates of the object. In object detection true positive is defined as prediction with predicted category equal to the ground truth label and Intersection Over Union (IOU) above the predefined threshold, usually $0.5$. Successful prediction meets both criteria, otherwise, it is considered as a missed detection. In our case we substitute the IoU requirement with the distance metric $d=\sqrt{\Delta\phi^2+\Delta\eta^2}<33$~pixels as we regress only the centre of the box and box dimensions are universal across target classes. 

Our investigation into inference does not find any systematic issues. Occlusion, such as the one in $t \rightarrow bW$ decay, where jets are near, is not an obstacle against correct detection. Also, the jets close to the image edges are, generally, correctly classified.

To evaluate the model we use precision (or positive predictive value, PPV, $\frac{TP}{TP+FP}$) and recall (true positive rate, TPR) curve, and an average precision metric (AP), see Figure~\ref{Figure:Precision-Recall-Curve}. Intuitively, precision measures how accurate the predictions are while recall measures the quality of the positive predictions. Collectively, they determine how well the found set of jets corresponds to the set we expect to find. To draw a precision-recall (PR) curve, the predictions are first sorted in order of confidence followed by calculation of PPVs and TPRs for each confidence threshold. We held out $90$k samples as our test dataset. The TWN network results are closely matching the results of the FPN. TWN benefits from the long retraining period, as it yields marginally better AP. For performance details across target jet classes see Table~\ref{Table:Precision-Recall-Details}.

Throughout, we compare PFJet-SSD to the {\em baseline} which is a physics-based algorithm combining a jet soft-drop mass~\cite{Larkoski:2014wba} selection (under a specific mass hypothesis) and N-subjettiness~\cite{Thaler:2010tr} (to increase tagging purity). In particular, we require $65 < m < 105$ for V jets, $105 < m < 140$ for H jets, $105 < m < 210$ for t jets.

This physics-motivated baseline has performance that is typical of a rule-based state-of-the-art substructure jet tagger, with the typical recall of 0.3 for the precision of 0.6.

Figure~\ref{Figure:Precision-Details} shows the dependence of the precision at fixed recall across different jet classes. The precision is rather flat in all cases. The TWN results match closely the FPN ones, while an overall drop in performance (approximately constant across $\eta$, $\phi$, and $p_T$) is observed for the INT8 network. A drop is observed at the boundaries of the $\eta$ region, as a consequence of jets leaking out of acceptance at the edge of the endcaps (missing information of a part of the shower). Such a drop is not observed in the $\phi$ dimension suggesting that the network can handle the periodicity of the image. The precision across $p_T$ stays relatively flat, however, the sudden drop in the high $p_T$ region of V jets is due to the low number of samples in that region, see the details in Section~\ref{Section:Experiments}.

Figure~\ref{Figure:Displacement-Regression-Error} shows the residual in the determination of $\eta$ and $\phi$ and the ratio of the reconstructed-to-true jet $p_T$, as a function of the jet $p_T$ for the different classes.

\begin{figure}[ht]
  \centering
  \includegraphics[width=.6\linewidth]{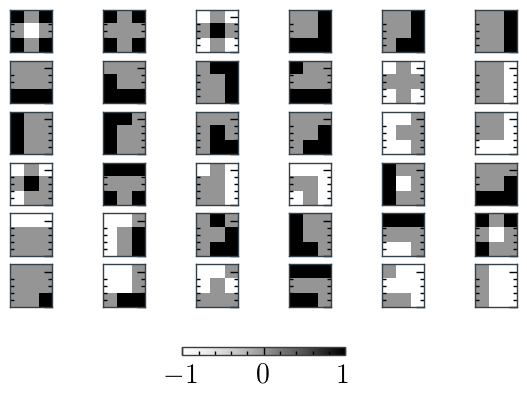}
  \caption{Most common filters of the PFJet-SSD TWN.}
  \label{Figure:TWN-Filters}
\end{figure}

Finally, we visualize the most repeating filters of the TWN in Figure~\ref{Figure:TWN-Filters}. Remarkably, the network optimizes to use a set very similar to the commonly used ones, e.g. smoothing, corner detection or edge detection filters.

\begin{figure}[ht]
  \centering
  \includegraphics[width=\linewidth]{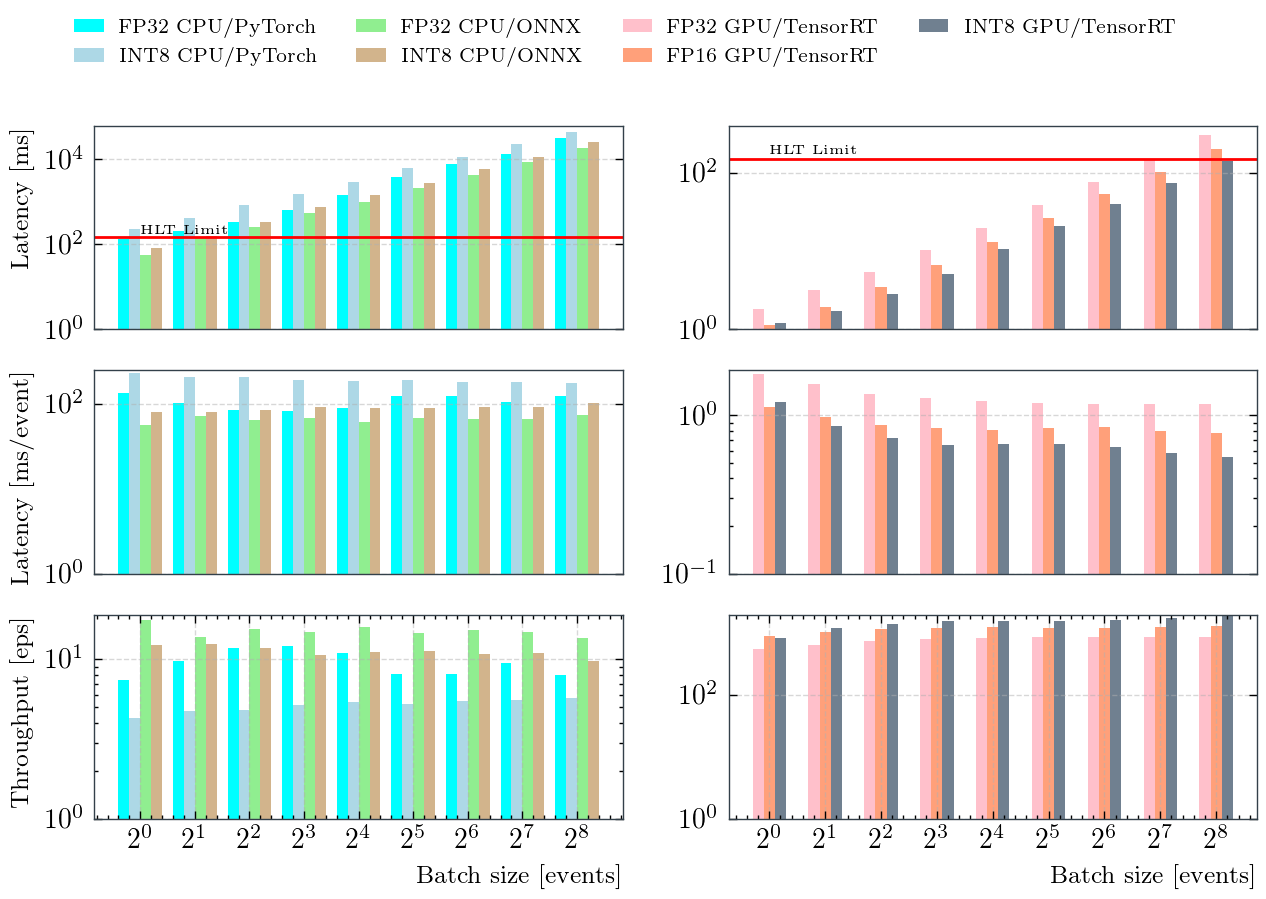}
  \caption{Comparison of inference latency and throughput for different versions of PFJet-SSD running on different platforms.}
  \label{Figure:Inference-Results}
\end{figure}

\paragraph{Latency and Power Measurements}
We investigate the latency and throughput of the proposed algorithm on architectures where parallel computing is more adequate. We compare the baseline, running native PyTorch inference on the Intel Xeon Silver 4114 CPU with ONNX accelerated version and TensorRT optimized version on Nvidia Tesla V100. Results are given in Fig~\ref{Figure:Inference-Results}, separately for CPUs and GPUs. Having in mind an offline application, one could maximally benefit if the network throughput by running the network at once across batches of events, e.g., implementing the inference-as-a-service concept discussed in~\cite{Krupa:2020bwg}.  

While the inference-as-a-service paradigm could also be implemented online, the current design of HLT farms foresees that processing parallelization is achieved by sending different events to different computing units. In this context, the batch size is constrained to one, since the inference of the proposed SSD model happens per event. In this case, execution on CPU would be borderline, within the average event processing latency but consuming most of it. On the other hand, moving the execution to a GPU would reduce the execution time to negligible levels. This could be particularly interesting under the assumption that GPUs would be used to run the local reconstruction~\cite{Bocci:2020pmi,Rovere:2020rqi,Qasim:2021hex} and the creation of PF candidates~\cite{Pata:2021oez}. 

Deep learning inference at scale requires high power consumption, especially with the use of GPUs and CPUs. It is possible to keep the power and die area, down by using an AI-specific hardware platform as is used in edge devices. Since edge devices usually operate on batteries where power is a limited resource, AI-specific hardware platforms for edge devices are highly power efficient. With smaller die areas, manufacturing costs and power consumption can be reduced.

SensPro is a family of ultra-light AI DSPs that can perform efficient inference while consuming only a fraction of the power and area used by GPUs and CPUs. CEVA's hardware platform for jet detection consists of a stack of ten SensPro (SP) DSP cores. Each core delivers 2 TOPS. An additional SP core is added to serve as a controller. This solution delivers 20 TOPS and can run TWN natively, reaching latency comparable to a GPU running an 8-bit network. This proposed layout has orders of magnitude lower area and power consumption than GPU and CPU, see Table~\ref{Table:Die-Are-Power-Latency}. The SP ultra-light solution can also be synthesized to an FPGA and used in collision detection. 

\begin{table}[ht]
    \centering
    \begin{footnotesize}
    \begin{tabular}{|l|c|c|c|} \hline
         & Die Area (mm2) & Power (W) & Latency (ms) \\ \hline
         DSP CEVA SP1000 2x8 & 0.77 & 0.75 & 8.5 \\ \hline
         DSP CEVA 10xSP1000 + Controller 2x8 & 8.47 & 8.25 & 0.9 \\ \hline
         GPU Nvidia Tesla V100 8x8 & 815 & 250 & 1.1 \\ \hline
         CPU Intel Xeon Silver 4114. 32float & 4,294 & 85 & 134 \\ \hline
    \end{tabular}
    \end{footnotesize}
    \caption{Die area, power and latency measurements for different hardware architectures. The latency is measured for inference on a single input. The reason for this is that collision detection is done sequentially in real-time. The input data is fed to the network directly from the sensor without storing it.}
    \label{Table:Die-Are-Power-Latency}
\end{table}
\section{Conclusions}\label{Section:Conclusions}
We propose a fast and lightweight detection algorithm for jet tagging and reconstruction based on computer vision techniques. Naturally high precision and generalization are required, but nuisance factors of variations can break the algorithm. That makes this problem hard. Intra-class variations, such as perspective distortion, e.g. rotation; densely arranged jets (occlusion); or blurred signatures (the detector response may not be clear) are common challenges. Besides, jets are small objects, a reappearing issue with object detectors and background pileup may further disturb their visual appearance. Thus, robustness to detector effects, its imperfections and failures is required. 

Even after a successful proof-of-concept deployment to production will still produce challenges as many of the problems lay outside of the simulation. More importantly, the real-time detection requirements force further investigations into more optimizations on algorithm and hardware runtime.

The PFJet-SSD paves the way for solving these issues. The algorithm did not experience accuracy drops during pruning, suggesting that the depth of the network is more important than the width. The number of channels can likely be reduced further and thus speed up computations. We observed a gap between TWN and INT8 performance which suggests to us that the optimal quantization level could be achieved through mixed-precision, a possible direction for future studies.

From the physics point of view, the algorithm manifests a promising behaviour in low momentum regions out of reach for the baseline model, see high precision results in Figure~\ref{Figure:Precision-Details}. This makes it interesting for studies on Vector Boson Fusion jets but the performance against QCD needs to be measured.

\section*{Acknowledgments}\label{Section:Acknowledgments}
We thank Loukas Gouskos and Huilin Qu for useful discussions and suggestions. A.~A.~P., M.~P., S.~S. and V.~L. are supported by the European Research Council (ERC) under the European Union's Horizon 2020 research and innovation program (grant agreement n$^o$ 772369). A.~A.~P. is supported by CEVA under the CERN Knowledge Transfer Group.

\clearpage
\newpage
\bibliographystyle{iopart-num}
\bibliography{biblio}
\end{document}